\documentclass[sigconf]{acmart}

\AtBeginDocument{%
  \providecommand\BibTeX{{%
    \normalfont B\kern-0.5em{\scshape i\kern-0.25em b}\kern-0.8em\TeX}}}

\copyrightyear{2026}
\acmYear{2026}
\setcopyright{cc}
\setcctype{by-nc-nd}
\acmConference[CHI '26]{Proceedings of the 2026 CHI Conference on Human Factors in Computing Systems}{April 13--17, 2026}{Barcelona, Spain}
\acmBooktitle{Proceedings of the 2026 CHI Conference on Human Factors in Computing Systems (CHI '26), April 13--17, 2026, Barcelona, Spain}
\acmPrice{}
\acmDOI{10.1145/3772318.3791895}
\acmISBN{979-8-4007-2278-3/2026/04}

\usepackage{colortbl}
\usepackage{xcolor}
\usepackage{xspace}
\usepackage{stfloats}

\newcolumntype{L}{>{\centering\arraybackslash}X}
\newcommand{\tangiblesite}[0]{\textsc{TangibleSite}\xspace}

\begin{document}

\title{As Content and Layout Co-Evolve: \tangiblesite for Scaffolding Blind People's Webpage Design through Multimodal Interaction}

\author{Jiasheng Li}
\email{jsli@umd.edu}
\affiliation{%
  \institution{University of Maryland}
  \city{College Park}
  \state{Maryland}
  \country{USA}
  \postcode{20742}
}
\authornotemark[1]

\author{Zining Zhang}
\email{znzhang@umd.edu}
\affiliation{%
  \institution{University of Maryland}
  \city{College Park}
  \state{Maryland}
  \country{USA}
  \postcode{20742}
}
\authornote{Co-first authors, equal contribution.}

\author{Zeyu Yan}
\email{zeyuy@umd.edu}
\affiliation{%
  \institution{University of Maryland}
  \city{College Park}
  \state{Maryland}
  \country{USA}
  \postcode{20742}
}

\author{Matthew Wong}
\email{mwong119@terpmail.umd.edu }
\affiliation{%
  \institution{University of Maryland}
  \city{College Park}
  \state{Maryland}
  \country{USA}
  \postcode{20742}
}

\author{Arnav Mittal}
\email{amittal5@terpmail.umd.edu}
\affiliation{%
  \institution{University of Maryland}
  \city{College Park}
  \state{Maryland}
  \country{USA}
  \postcode{20742}
}

\author{Ge Gao}
\email{gegao@umd.edu}
\affiliation{%
  \institution{University of Maryland}
  \city{College Park}
  \state{Maryland}
  \country{USA}
  \postcode{20742}
}
\authornotemark[2]

\author{Huaishu Peng}
\email{huaishu@umd.edu}
\affiliation{%
  \institution{University of Maryland}
  \city{College Park}
  \state{Maryland}
  \country{USA}
  \postcode{20742}
}
\authornote{Co-last authors, equal contribution.}

\renewcommand{\shortauthors}{Li, et al.}

\begin{abstract}
Creating webpages requires generating content and arranging layout while iteratively refining both to achieve a coherent design, a process that can be challenging for blind individuals. To understand how blind designers navigate this process, we conducted two rounds of co-design sessions with blind participants, using design probes to elicit their strategies and support needs. Our findings reveal a preference for content and layout to co-evolve, but this process requires external support through cues that situate local elements within the broader page structure as well as multimodal interactions. Building on these insights, we developed \tangiblesite, an accessible web design tool that provides real-time multimodal feedback through tangible, auditory, and speech-based interactions. \tangiblesite enables blind individuals to create, edit, and reposition webpage elements while integrating content and layout decisions. A formative evaluation with six blind participants demonstrated that \tangiblesite enabled independent webpage creation, supported refinement across content and layout, and reduced barriers to achieving visually consistent designs.

\end{abstract}



\begin{teaserfigure}
  \centering
  \includegraphics[width=\textwidth]{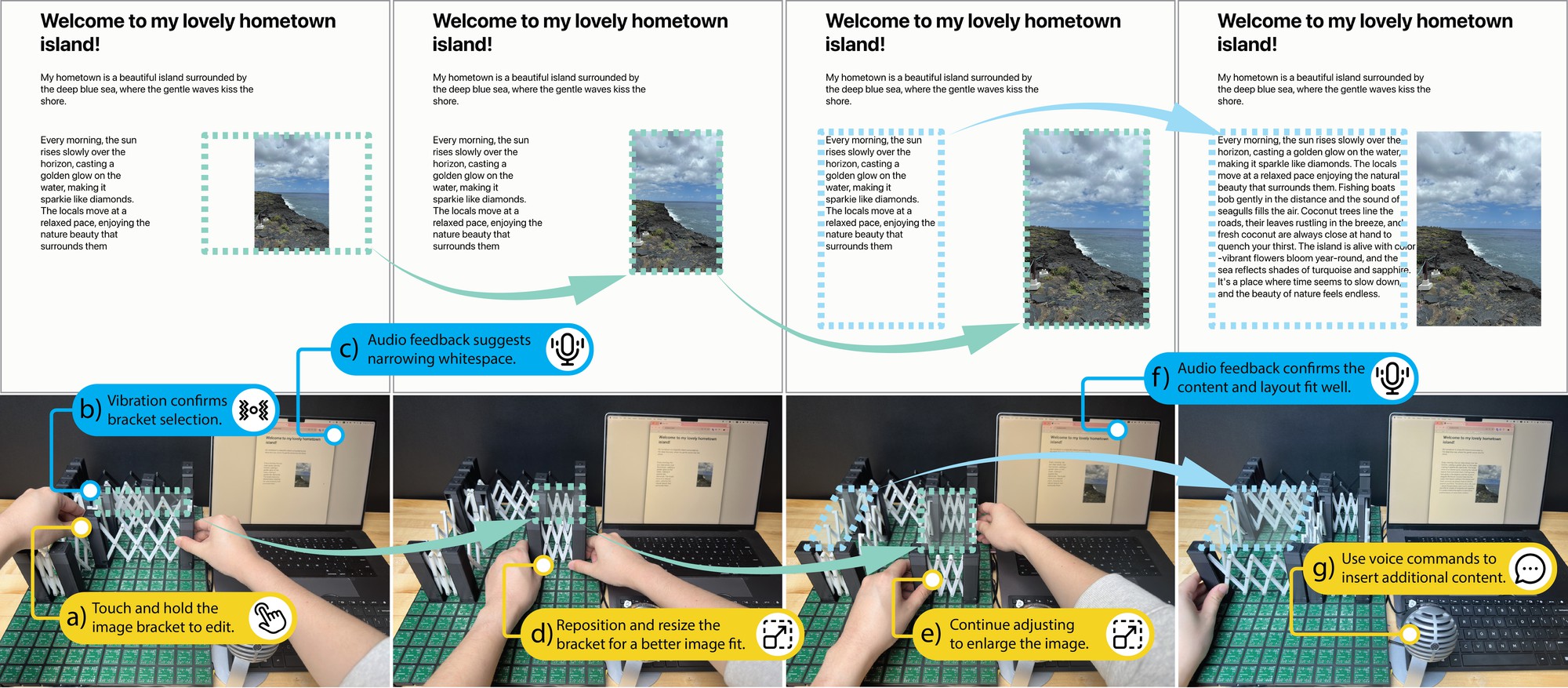}
  \caption{A blind individual iteratively designs a blog webpage using \tangiblesite. The top row shows the evolving webpage rendering, while the bottom row shows the corresponding actions performed by the blind designer. 
  The designer a) activates an existing image element (highlighted in mint green) by touching and holding the image bracket, which b) triggers editing mode with vibration feedback. c) Audio feedback detects whitespace on both sides, suggesting the bracket should be narrowed to reduce excess space. The designer d) repositions and resizes the image bracket on the baseboard.  The designer e) continues adjusting the image bracket to slightly enlarge the image until f) audio feedback confirms that it fits the layout. The designer then switches to text editing by activating the text bracket (highlighted in light blue) on the left, and g) uses a voice command to insert additional content.
  \Description{Four-panel figure showing \tangiblesite in front of a laptop with a synchronized webpage preview. Panel a: a blind individual performs a long touch (over 3 seconds) on the image bracket to activate it for editing and receives audio feedback indicating a mismatch between the bracket size and the image size. Panel b: the user adjusts the position and size of the image bracket to better fit the image. Panel c: the user enlarges the image by adjusting the image bracket on the baseboard and receives audio feedback confirming that the image now fits the bracket. Panel d: the user activates the text bracket by touching it for more than 3 seconds and uses a voice command to insert text.}
  }
  \label{fig:teaser}
\end{teaserfigure}

\maketitle

\section{Introduction}

Design is, by nature, a process of exploration and refinement. It requires evaluating the current state, identifying areas for improvement, making adjustments, and then reassessing the results. In web design, this often takes the form of continuously reviewing the visual arrangement of elements and considering how their layout and composition shapes the overall experience.

For sighted designers, the iterative cycle is greatly supported by rapid prototyping tools that provide instant visual feedback (e.g., Figma) or coding environments with built-in previews (e.g., Visual Studio Code). For blind individuals, however, perceiving a webpage, identifying differences, and applying changes is undeniably challenging. While screen readers can support experienced and highly-skilled blind designers in website creation~\cite{kearney_web_development, google_site_accessibility}, these tools convey limited spatial information about web elements and provide limited access to visual formatting~\cite{Understanding_zhang}.

In response, HCI researchers have been developing various assistive technologies to lower barriers to web design for blind individuals~\cite{potluri_web_design, ide_accessible_1, li_tangibleGrid}. For example, Li et al. introduce a system that uses embossed tactile templates overlaid on a tablet to help blind individuals explore and design spatial webpage layouts~\cite{jingyi_layout}. TangibleGrid~\cite{li_tangibleGrid} provides a tangible interface for real-time webpage layout editing. DesignChecker~\cite{designChecker}, a browser extension, enables blind web developers to compare their designs with online references and edit HTML or CSS files. While these systems advance support for creating and inspecting graphical web representations, they provide limited support for an end to end web design workflow in which a webpage can be created from scratch and iteratively refined through coherent editing of both content and layout.

In this work, we aim to fill this gap by enabling blind individuals to create and iteratively refine webpages through integrated editing of content and layout. To ground this goal in blind designers’ strategies and support needs, we first conducted two rounds of co-design sessions with three blind participants using a tangible design probe inspired by TangibleGrid~\cite{li_tangibleGrid}. 

In these sessions, we observed that participants preferred content and layout to co-evolve, refining both in tandem when possible. When designs were simple, they alternated naturally between editing content and rearranging layout elements (Figure~\ref{fig:c1p2_workflow}). As layouts became more complex, however, this flexibility diminished and participants expected more external cues to stay oriented (Figure~\ref{fig:c1p3_workflow}). This suggests a need for systems that preserve structure across edits so blind designers can verify changes without reconstructing the page from memory. We also observed that touch played a central role in inspecting elements, confirming system state, and planning next steps (Figure~\ref{fig:sweeping_board}). Because similar touch actions often served different functions, touch interactions should clearly distinguish intents such as inspecting, selecting, editing, and moving through touch coupled feedback.
Participants additionally expected explicit, actionable audio feedback on content fit and element relationships within the layout.

Based on these findings, we designed \tangiblesite, an accessible web authoring environment that supports webpage creation as content and layout co-evolve (Figure~\ref{fig:teaser}). To reduce blind designers’ memory burden, \tangiblesite maintains an up-to-date representation of the page’s structure and content. It provides real-time multimodal feedback at both the element and page levels, supporting refinement of individual elements and understanding of the overall relationship between content and layout. To evaluate \tangiblesite, we conducted a formative study with six blind participants. All participants created a webpage and iterated on both layout and content. They reported increased control over the design process and relatively low frustration during authoring.

In summary, our paper contributes: (1) an empirical investigation of blind designers’ preferred workflow for integrating content and layout in web design; (2) \tangiblesite, an accessible multimodal web authoring prototype that supports real-time feedback and iterative refinement of both; and (3) a formative evaluation with six blind participants demonstrating end-to-end~webpage~creation~and~refinement.

\section{Related Work}
\subsection{Nonvisual Access to Graphical Interfaces}
A substantial body of work has explored how audio and haptic modalities support blind people in navigating and understanding existing graphical interfaces.
Early research, such as Sonic Grid~\cite{SonicGrid_2008Jagdish}, uses auditory icons and speech feedback to support navigation within digital interfaces. 
Gesture based overlays that integrate with screen readers such as VoiceOver and JAWS provide additional ways for blind users to access on-screen elements ~\cite{sliderule_2008kane, accessOverlay_2011Kane}.

Although these approaches enable effective access to text, they offer limited support for understanding spatial or graphical structures. 
To address this gap, researchers have introduced accessibility software~\cite{imageexplorer_2022lee, voxlens_2022sharif,evographs_2018sharif}, sonifications~\cite{edgesonic_2011yoshida}, and auditory cueing strategies~\cite{design_2003brown, plumb_2006Calder, teaching_2006cohen, audiograf_1996kennel} that describe or encode visual information. 
ImageExplorer~\cite{imageexplorer_2022lee} uses a suite of deep learning models to generate multilevel image descriptions, while VoxLens~\cite{voxlens_2022sharif} supports querying graphs, tables, and charts.  
EdgeSonic~\cite{edgesonic_2011yoshida} conveys object boundaries on touchscreens by encoding distance to edges through pitch and timing. 
Additional work has extended nonvisual access to maps~\cite{geospatial_2022sharif,indoor_map_layout_calle2016, timbremap_2010su}. 
For example, Jimenez et al.~\cite{indoor_map_layout_calle2016} propose an SVG encoding of indoor layouts that enables screen readers to verbalize spatial structure.

In parallel, haptic and tangible interfaces have been explored to complement these auditory techniques.
Vibrotactile patterns have been used to encode geometric shapes~\cite{evaluation_2013awada, haptic_2018palani}.
Tactile graphics and refreshable tactile displays provide representations of textures, mathematical content and graphs~\cite{exploring_2014mullenbach, surface_2013mullenbach, tactile_2011xu, comparing_2019simonnet, wall2006feeling, automated_2007Jayant,automating_2005ladner, touchAndSpeech_1998blenkhorn, geographic_2010Delogu}.
Other work has introduced 3D printed tactile artifacts for maps~\cite{3Drinted_map_Crawford, 3D_printed_map_Holloway, accessible_map_3D_Holloway} and graphs~\cite{3Dprint_tactile_graphics_Clark}. 
TouchPlates~\cite{touchplates_2013kane}, for example, uses physical overlay on touchscreens to help blind people locate interactive virtual buttons.

Taken together, this body of work provides extensive solutions for interpreting existing visual interfaces. However, the focus has largely remained on access and interpretation rather than on supporting blind individuals as creators of visual or spatial content.

\subsection{Visual Content Authoring Tools for the Blind People}
Alongside work on nonvisual access, researchers have developed systems that support blind individuals as creators of visual media, as blind individuals have expressed strong interest in producing drawings, photographs, and other visual artifacts~\cite{Understanding_zhang}.
Early tools such as TDraw~\cite{tdraw_1996kurze} enable nonvisual drawing on swell paper with audio feedback. More recent drawing systems such as AudioDraw~\cite{audiodraw_2016grussenmeyer} and A11yBoard~\cite{a11yboard_googleside, zhang_a11yboard} use gestures, vibrotactile cues and auditory descriptions to help blind individuals construct geometric shapes on digital artboards. 
Tools such as VizSnap~\cite{vizsnap_2016adams} and EasySnap~\cite{EasySnap_2011jayant} support blind individuals in taking photographs by providing audio guidance for framing.

Recent progress in AI has broadened what blind creators can produce. 
A11yShape~\cite{a11yshape_2025zhang} assists blind users in creating 3D models. GenAssist~\cite{genassist_2023mina} provides guided image generation and refinement. AVScript~\cite{avscript_2023huh} supports nonvisual editing of video clips. 
AltCanvas~\cite{altcanvas_2024lee} decomposes an image into tiles and uses AI to generate or edit each tile before recombining them into a complete image. 
Across these systems, language and vision models help blind individuals describe, generate, and edit visual media with control.

Blind creators also face barriers in evaluating visual outcomes or understanding visual concepts, and therefore frequently rely on sighted collaborators to assess their designs~\cite{Understanding_zhang}. Diffscriber~\cite{diffscriber_2022peng} helps blind slides creators understand changes made by sighted collaborators to both slide content and layout by automatically identifying and describing those edits. To further support blind creators in assessing their own slide designs, Zhang et al.~\cite{slideaudit_2025zhang} build a large dataset to support automated slide evaluation. DesignChecker~\cite{designChecker} analyzes webpages created by blind programmers and provides code-level suggestions to refine layout and content. VizXpress~\cite{vizxpress_2025zhang} assesses photographic aesthetics and provides editing guidance.

Despite these efforts to support blind people in drawing and visual content editing, most systems operate within a single unit of content (e.g., one image, one photo) rather than addressing multi-element graphical designs. In this paper, we instead focus on entire webpages composed of multiple elements such as text paragraphs, images, and videos, and explicitly investigate how to help blind designers identify and understand content–layout issues. \tangiblesite gives blind designers control over the spatial organization of these elements, not just the content itself.

\begin{table*}[t]
  \centering
  \small
  \caption{Comparison of systems that support accessible webpage design.}
  \label{tab:design-support}
  \begin{tabular*}{\textwidth}{@{\extracolsep{\fill}} l c c c c @{}}
    \toprule
    \textbf{System} &
    \textbf{\shortstack{Tangible Layout\\Editing}} &
    \textbf{\shortstack{Content Creation\\and Editing}} &
    \textbf{\shortstack{Real Time Feedback\\for Iteration}} &
    \textbf{\shortstack{Content--Layout Fit\\Guidance}} \\
    \midrule
    Potluri et al.~\cite{potluri_web_design} & \(\times\) & \(\triangle\) & \(\times\) & \(\times\) \\
    Li et al.~\cite{jingyi_layout}          & \(\checkmark\) & \(\times\) & \(\times\) & \(\times\) \\
    TangibleGrid~\cite{li_tangibleGrid}     & \(\checkmark\) & \(\times\) & \(\triangle\) & \(\times\) \\
    \textbf{\tangiblesite (ours)}            & \(\checkmark\) & \(\checkmark\) & \(\checkmark\) & \(\checkmark\) \\
    \bottomrule
  \end{tabular*}
\end{table*}

\subsection{Accessible Web Creation}
Creating webpages traditionally requires knowledge of HTML, CSS, and JavaScript. Considerable effort has thus been devoted to better equipping blind individuals interested in web programming through workshops, bootcamps ~\cite{bootcamp_kearney, workshop_kearney, workshop_kearney_2, kearney_web_development}, and online courses~\cite{cisco_edu, cucat}. Yet, programming tools themselves often remain inaccessible~\cite{blind_coding_challenges_1, blind_coding_challenges_2, blind_coding_challenges_3, blind_coding_challenges_4}. 
For example, screen readers do not fully convey code structure, leading to over reliance on memory~\cite{blind_coding_challenges_5}. 
To address these issues, researchers have developed various tools to improve code understanding~\cite{ide_accessible_3, ide_accessible_4, data_structure_for_blind}, navigating in-line locations~\cite{ide_accessible_2, ide_accessible_6}, and debugging~\cite{ide_accessible_1, ide_accessible_5, nvda_addon}. While effective for developers who are comfortable with code, these approaches remain challenging for many novices or blind individuals who prefer non code based design workflows.

The primary approach to codeless web design is through graphical web authoring tools such as Figma, WordPress, and Google Sites. %
More recently, advances in generative AI have enabled intent-driven and example-driven authoring workflows~\cite{uicoder_2024wu, codedesign_2025, de-stijl_2023shi, guicomp_2020lee,rewire_2018Swearngin, design2code_2024si}.
One such example is MISTY, a workflow that leverages LLMs to help designers rapidly retrieve online examples within ongoing designs, accelerating exploration and iteration~\cite{misty_2025lu}. 

Despite this progress, most graphical web design tools, whether traditional or more recent, are not designed with blind individuals in mind, and few offer nonvisual feedback. Complex layout information combined with page content is difficult to convey through standard screen readers. 

To lower barriers to web design for blind individuals, recent HCI research has explored nonvisual design approaches such as tangible representations, tactile printouts, and touchscreen-based design systems. 
Li et al.~\cite{jingyi_layout} introduce a system that prints tactile webpage templates to overlay on a touchscreen, enabling blind individuals to explore page structure and perform layout edits through touch and audio. While effective for conveying spatial relationships, the need to regenerate and reprint tactile sheets after each edit slows the feedback cycle and can make refinement challenging for complex or frequently changing designs.
TangibleGrid~\cite{li_tangibleGrid, tangiblegrid_2} introduces a physical baseboard with shape-adjustable brackets to represent layouts, enabling direct creation, adjustment, and deletion of elements with real time feedback. Although intuitive and spatially grounded, it focuses solely on layout structure and does not support content editing, which can make full webpage authoring less seamless. 
Other systems, such as the one introduced by Potluri et al.~\cite{potluri_web_design}, use touchscreens to edit element attributes (e.g., padding, font size) and automatically update CSS source files. This approach provides control over styling, but its emphasis on attribute level adjustments offers less support for understanding or manipulating the larger layout structure of a webpage. Across these efforts, the focus has largely remained on layout representation and manipulation, with few tools supporting content entry alongside layout design or providing real time feedback for both during design.

Our \tangiblesite builds on this prior work by enabling blind individuals to create webpages through multimodal interaction. Unlike previous systems, \tangiblesite supports content and layout design as they co-evolve by coupling tangible layout manipulation with audio feedback. Tangible interaction provides a spatial scaffold for externalizing and revisiting layout decisions, while audio feedback evaluates content fit, signals layout conflicts, and suggests concrete adjustments. Table~\ref{tab:design-support} summarizes the key differences between \tangiblesite and previous systems.

\section{Understanding Blind Individuals' Interactive Workflows}
Designing a webpage involves coordinating content and layout, often in ways that depend on nuanced visual cues such as proportion and spatial alignment. For blind individuals, the absence of these visual affordances makes the process significantly more challenging, especially when it involves multiple elements and and when iterating between local and global decisions.
To investigate how blind individuals conceptualize and navigate content-layout relationships, we conducted two rounds of co-design sessions using a tangible design probe modeled after TangibleGrid. Our aim was to surface both the challenges blind individuals face and the interaction mechanisms needed to support them. Our study was guided by the following research questions:

\textbf{RQ1.} How do blind individuals conceptualize and organize the workflow of content and layout during webpage creation?

\textbf{RQ2.} What types of information help blind individuals detect conflicts between local content and layout, and maintain page-level consistency during iteration?

\textbf{RQ3.} What features and mechanisms should authoring tools provide to help blind individuals resolve global layout consequences of local edits?

These questions shaped the design of our co-design sessions and served as the analytic structure for the findings below.

\subsection{Apparatus}\label{apparatus}
We used a simplified version of TangibleGrid as our primary design probe.
The probe consisted of a 3D-printed tactile baseboard whose groove structure represented the underlying grid of a webpage, paired with a set of resizable brackets representing web elements such as text blocks, images and videos. Brackets included tactile patterns indicating element types and could be resized by repositioning their corner pillars. This allowed participants to physically inspect size, location, and spatial relationships.

Unlike the full TangibleGrid system, our probe did not include audio or computation. Instead, one researcher acted as the system, interpreting participants' spoken commands, simulating audio feedback, and narrating digital changes. This low-fidelity approach ensured that participants’ strategies emerged naturally without being shaped prematurely by system assumptions. A keyboard was provided for participants who preferred typing over voice input.

\subsection{Participants}\label{participant_background}

We recruited three participants through the local National Federation of the Blind (NFB) email list~\cite{nfb}, and all completed both co-design sessions. Participants self-reported being blind and self-identified as women (ages 19--69, \textit{M} = 43, \textit{SD} = 20.4).

The purpose of the co-design sessions was to surface design practices that benefit from deep, contextual observation. We therefore worked closely with a small number of participants in extended sessions that involved layout construction, content entry, and iterative refinement with the tangible probe. This format supported close observation of participants’ choices and strategies and produced detailed behavioral and decision-making data. We summarize participants’ prior design experience below.

P1 had no prior experience with webpage design but had created 2D graphical content such as PowerPoint presentations. P1 typically used predefined templates for text and layout, rarely added images or modified visual appearance, and consistently relied on assistance from a sighted collaborator to verify the design.

P2 was actively involved in a project to develop a website focused on aging and disability consulting. P2 was responsible for gathering the necessary content for the site, but did not directly implement the webpages herself. Instead, P2 collaborated with a sighted partner who handled the web design tasks, including layout design and translating P2’s ideas into the final implementation.

P3 had experience developing websites in a Linux environment. P3 noted that the process required extensive memorization, particularly of coding syntax. Despite the availability of accessible tools for blind developers, P3 still faced considerable challenges, such as limitations in the comprehension of webpage layouts and the adjustment of web elements' dimensions.

This study received approval from the university’s Institutional Review Board (IRB), and each participant was compensated at a rate of \$35 per hour.

\begin{figure*}[t]
  \centering
  \includegraphics[width=\textwidth]{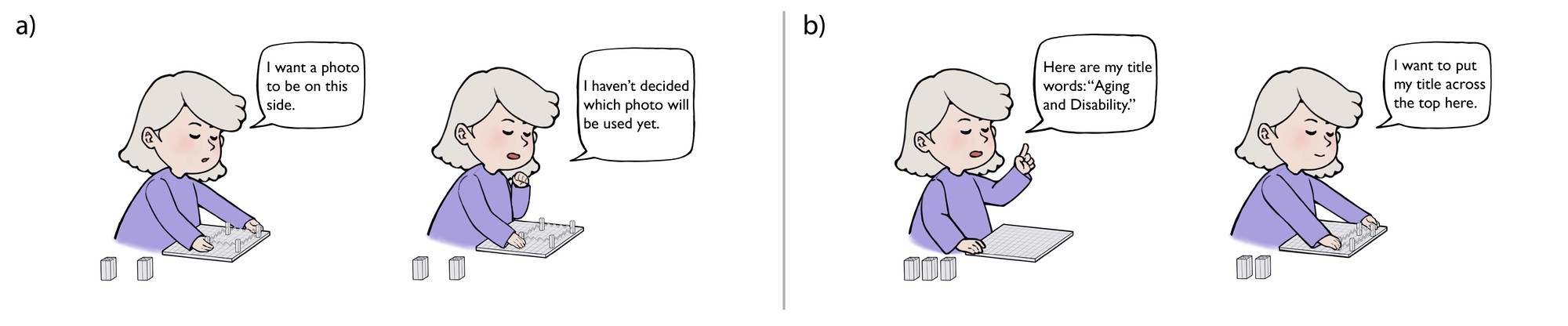}
  \caption{P2 demonstrated two different workflows. a) Placed the bracket first to establish the layout, then inserted the content. b) Spoke the content first, then placed the bracket to determine its placement.}
  \Description{This figure contains two sub-figures and it shows a different web design workflow from P2. a) Placing the bracket to design the layout first, then inserting the content, b) Speaking out the content first, then placing the bracket to decide where to place it.}
  \label{fig:c1p2_workflow}
\end{figure*}

\subsection{Data Analysis}
All sessions were video and audio recorded with consent.
We analyzed the data using a qualitative approach that combined video interaction analysis \cite{jordan1995interaction} and thematic analysis \cite{braun2006using}.
We first transcribed all spoken dialogue, including participants’ questions, reasoning, and commands, as well as the simulated system responses.
In parallel, we reviewed the video recordings to document nonverbal behaviors such as hand movements, interaction sequences with the tangible probe, and strategies not evident from speech alone.
We then conducted thematic analysis, iteratively coding transcripts and video notes to identify patterns in how participants conceptualized content and layout, detected conflicts, and responded to feedback.
Codes were refined through repeated comparisons across participants and between co-design rounds, resulting in a set of consolidated themes that informed our design implications.

\subsection{Co-design Session 1}
\subsubsection{Procedure}
After a brief interview about participants’ past experiences, we introduced our design probe and provided two design scenarios. Scenario 1 focused on a single-element webpage design.
Participants were asked to create one webpage element of their choosing using the probe. This scenario isolated basic operations and allowed us to observe how blind participants approached layout and content when no constraints were imposed. 
Scenario 2 involved designing within a more complex, multi-element layout.
Participants were asked to add new content or modify existing elements within a crowded spatial arrangement involving multiple brackets. This scenario reflected more realistic design challenges in which multiple components interact and spatial constraints matter.

At the end of the session, we conducted a semi-structured interview with open-ended questions focusing on participants’ design rationales, strategies for navigating and interpreting multiple elements, and challenges encountered during content creation and layout editing.

\subsubsection{Findings}
We report three findings (F1–F3) across two co-design sessions. This section presents only F1 from Co-design Session~1; Co-design Session~2 reports F2 and F3.

\textbf{F1. Participants used flexible workflows in simple designs, but this flexibility diminished as complexity increased.}
Co-design Session 1 shows that blind participants value flexible workflows during webpage creation, readily interweaving content creation and layout composition when the design is simple. However, once multiple elements are involved, this flexibility becomes much harder to sustain, and participants shift toward a layout-first strategy to manage spatial complexity.

In the single-element scenario, participants demonstrated considerable freedom in how they initiated design. Some began by placing a bracket first to anchor a spatial position, then described the content they wanted to associate with it (Figure~\ref{fig:c1p2_workflow}a). At other times, the same participant first thought through and verbalized the content, then placed the matching bracket afterward (Figure~\ref{fig:c1p2_workflow}b). In this single-element context, participants were able to iteratively refine their content.
For example, during one design iteration, P1 expanded a short text description by enlarging the corresponding bracket to create space for additional content. P1 highlighted the flexibility to adjust both structure and content:

\begin{quote}
    \textit{``Once you broaden the bracket, you can go back to adding content. That’s a good thing, because sometimes you get an idea, and then you can change the bracket and go back. I feel like that’s what I can do with the baseboard and brackets.''}--P1
\end{quote}

\begin{figure}[t]
  \centering
  \includegraphics[width=\columnwidth]{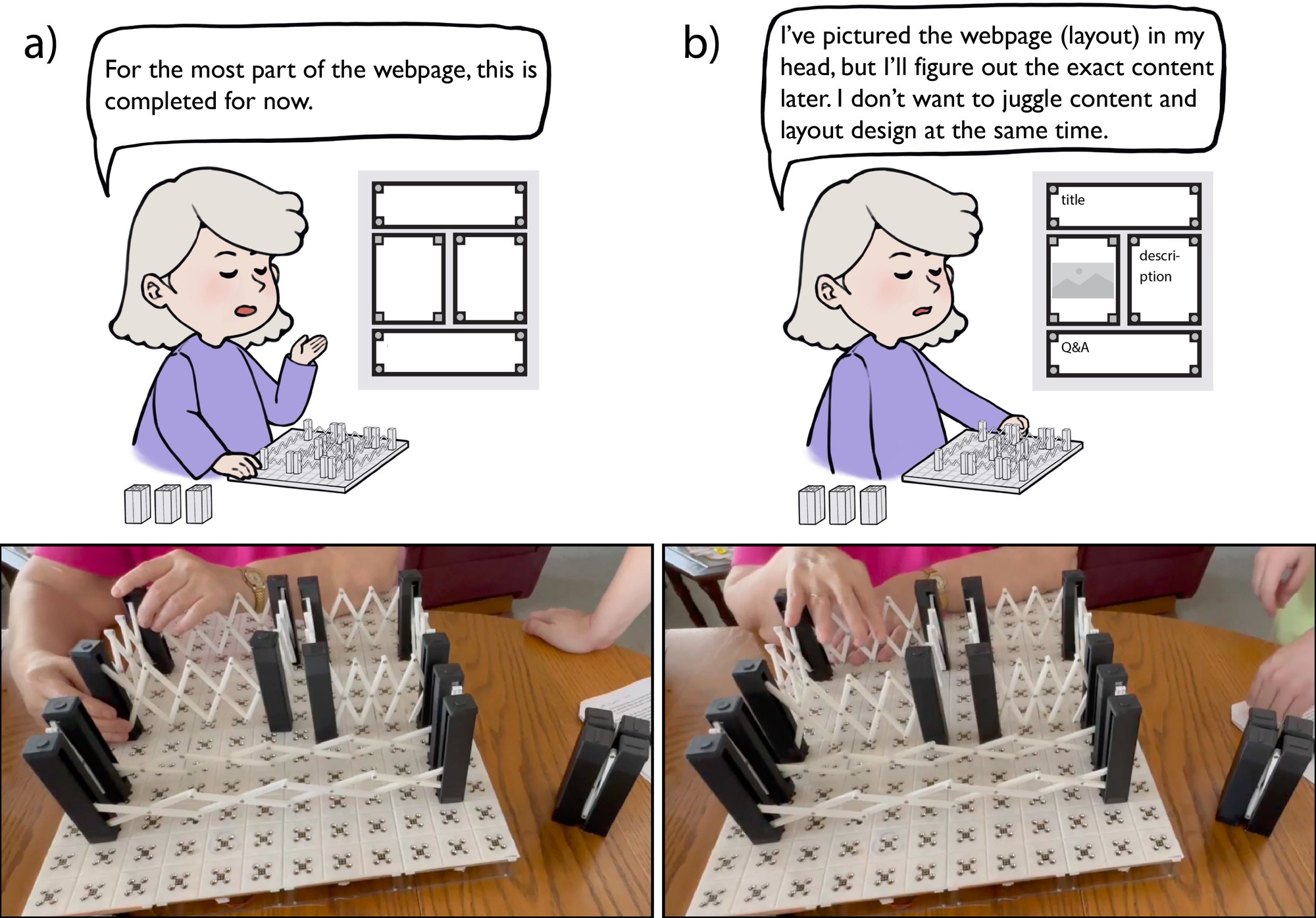}
  \caption{a) P2 used multiple brackets as placeholders to sketch the overall webpage structure. b) P2 then discussed the intended content for each bracket.}
  \Description{This figure contains two sub-figures. a) Participants placed multiple brackets on the baseboard as placeholders. b) Participant discussed their proposed content for each bracket.}
  \label{fig:c1p2_workflow_2}
\end{figure}
 
\begin{figure*}[b]
  \centering
  \includegraphics[width=\textwidth]{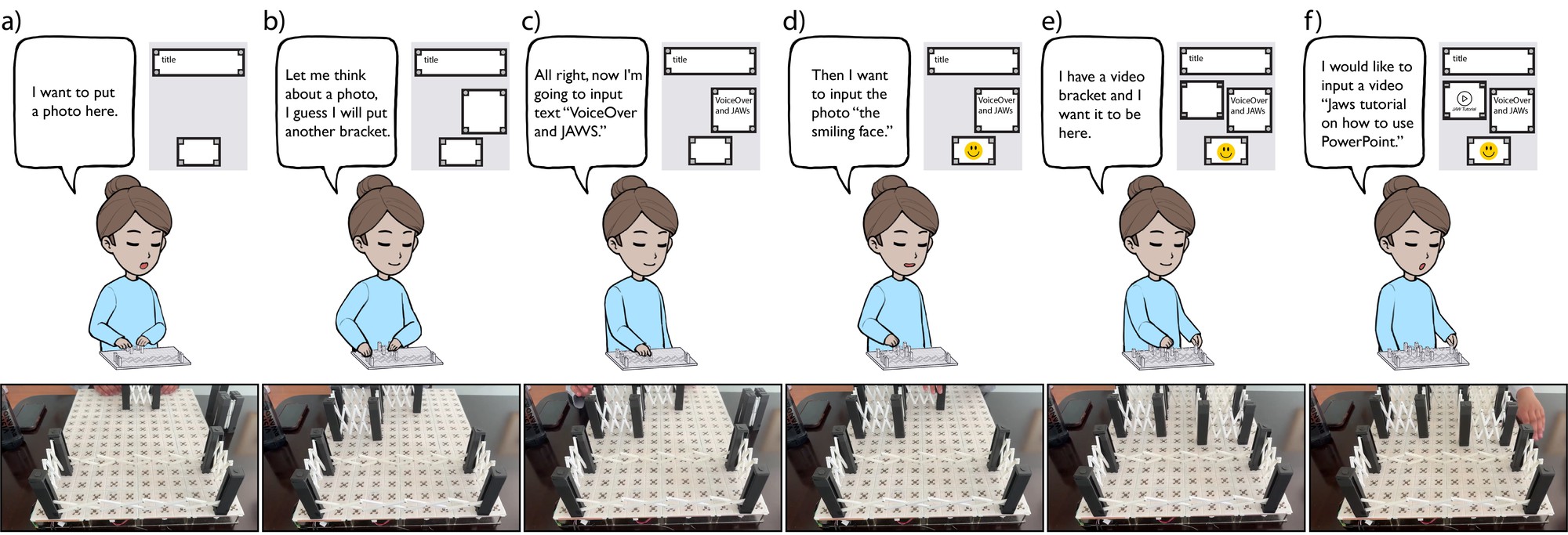}
  \caption{Workflow of P3 designing webpages using multiple web elements. a-c) P3 placed three web elements in sequence. d) P3 explored the board to verify the alignment and then inserted content into each bracket. e-f) P3 added an additional bracket on the baseboard and input its content.}
  \Description{This image contains 6 small figures showing the workflow of blind people design the web pages. Participants placed three web elements in sequence and explored the board to confirm the alignment, then insert content to each bracket. Later, added another bracket on the baseboard and input its content.}
  \label{fig:c1p3_workflow}
\end{figure*}

This flexibility, however, diminished as the webpage grew more complex. When multiple elements were involved, as in the scenario 2, participants no longer began with content. Instead, they prioritized establishing and verifying the layout. They spent time touching existing brackets to understand the current structure, then placed new brackets to secure the position and size before entering content. 
For example, in one session, P2 laid out four brackets in advance to establish locations for the elements they envisioned (a title, two images and a paragraph) and then returned to fill in the content for each (Figure~\ref{fig:c1p2_workflow_2}). Similarly, P3 placed brackets for three elements in sequence, explored the board again to confirm spatial relationships, and only then added content (Figure~\ref{fig:c1p3_workflow}a-d). P3 later added the final bracket in the same manner, establishing the layout before inserting the media content (Figure~\ref{fig:c1p3_workflow}e-f).

We also observed fewer edits in complex scenarios. Participants continued adding new elements but rarely revisited or modified earlier ones. As the number of elements increased, changing either content placement or layout relationships demanded re-establishing the mental model of the entire page, which became cognitively taxing. This applied to both spatial adjustments and decisions about how newly added content should relate to existing elements.

Overall, these findings suggest that nonvisual webpage design imposes growing cognitive demands as complexity increases. Therefore, rather than supporting only layout or only content, systems should help blind designers maintain a clear understanding of both the evolving content and the page's structure. Tools should help offload memory, provide summaries of what is present, highlight relationships among elements, and offer timely feedback that reduces the cost of reorientation. This raises the question of which types of feedback best support blind individuals in managing both content and layout as designs grow in complexity, which in turn motivated the focus of our Co-design Session 2.

\subsection{Co-design Session 2}
\begin{figure}[t]
    \centering
    \includegraphics[width=\columnwidth]{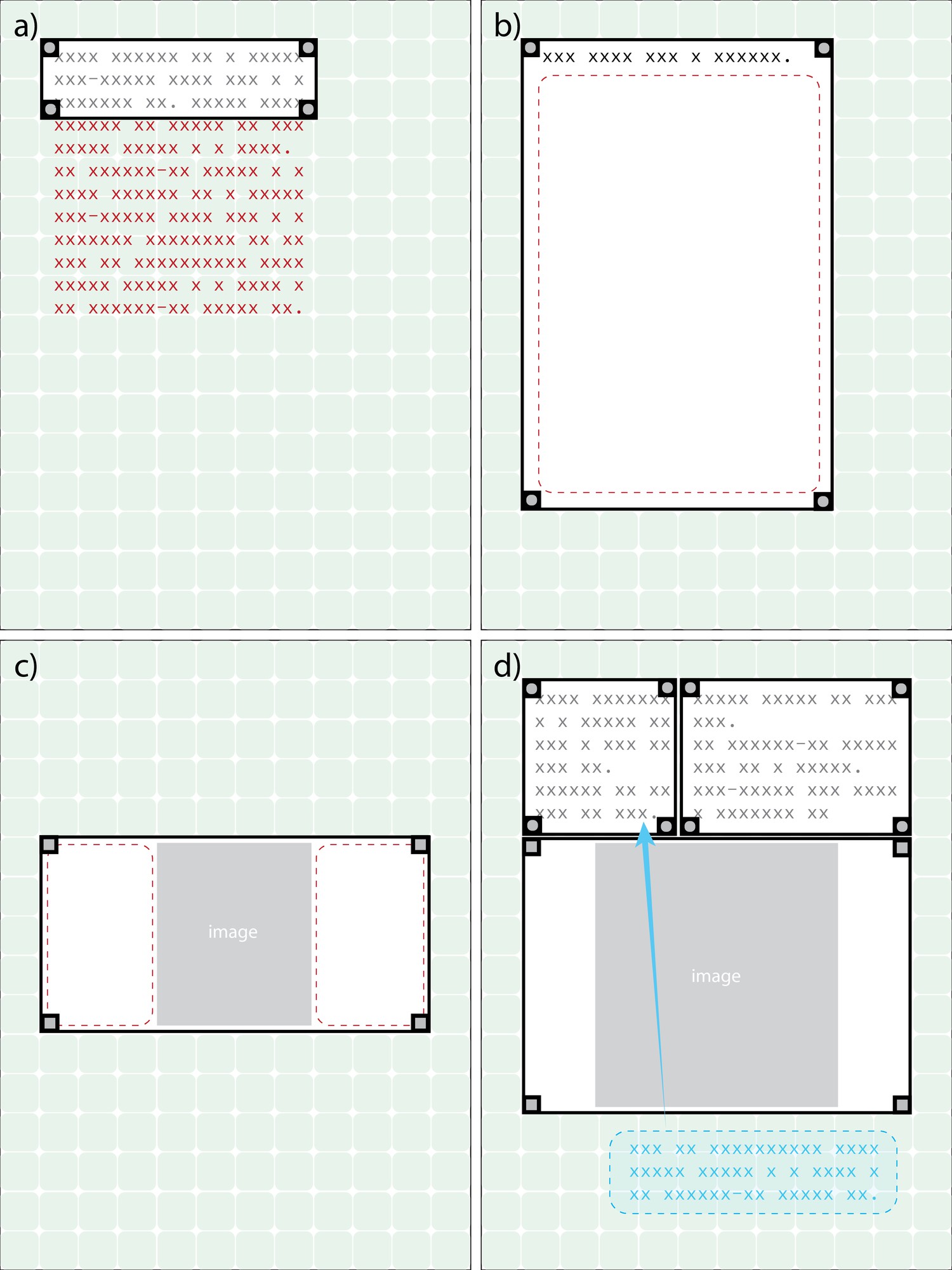}
    \caption{Four scenarios examined in Co-design Session 2: a) Text overflow: the text container is too small, causing content to extend beyond its boundary. b) Text underflow: the container is larger than the text, resulting in excessive unused space; the dashed inset marks the usable text area. c) Image–container mismatch: a landscape-oriented container holds a portrait image, leaving whitespace on the left and right. d) Multi-element reorganization: when an existing text container is full, adding new content requires resizing or repositioning neighboring elements to create space.}
    \Description{Four schematic panels on a square tactile grid, each with a black rectangular frame anchored at the corners. (a) A short, wide text region with rows of “x” characters spilling below the frame, indicating overflow. (b) A tall frame with a dashed inner boundary marking the usable text area; only a short line of placeholder text sits at the top, indicating underfill. (c) A wide frame split into three columns: a gray center block labeled “image” with empty side columns outlined by dashed rounded rectangles, showing a portrait image in a landscape container. (d) Two small text frames side-by-side above a large gray block labeled “image”; a dashed box below marks a caption area, indicating that adding more text will require reorganizing neighboring containers.}
    \label{fig:codesign2_scenarios}
\end{figure}

Building on insights from Co-design Session 1, 
we investigated how targeted system feedback could help blind individuals detect and negotiate content–layout conflicts. We presented participants with four common types of misalignment that occurred across text, image, and multi-element configurations (Figure ~\ref{fig:codesign2_scenarios}) and examined (1) what feedback was most helpful for participants in identifying these issues, (2) when such feedback should be delivered, and (3) how participants preferred to respond once a conflict was identified. 
Our broader aim was to understand how multimodal feedback can support blind designers in perceiving the current design state and maintaining an accurate mental model of the webpage.

\subsubsection{Design scenarios}
We introduced four scenarios that exemplified typical challenges blind individuals face when composing webpages using bounded containers.

The first two scenarios focused on text editing. Because webpage text must fit within bounded regions, participants encountered two opposite challenges: text overflow, where content extended beyond the container’s bounds (Figure~\ref{fig:codesign2_scenarios}a), and text underflow, where the container was much larger than the content, leaving substantial whitespace (Figure~\ref{fig:codesign2_scenarios}b). These issues are easy for sighted designers to notice at a glance but are much harder to perceive nonvisually.

The third scenario focused on image–container mismatch (Figure~\ref{fig:codesign2_scenarios}c). Even when inserting an image was straightforward, blind individuals could struggle to assess whether the container’s shape aligned with the image’s aspect ratio and whether the result involved unintended cropping or excess unused whitespace.

The fourth scenario addressed multi-element adjustment (Figure~\ref{fig:codesign2_scenarios}d). When attempting to add text to a filled container, blind individuals may find that neighboring elements constrained further expansion. This introduced a need to reposition or resize multiple elements simultaneously, which proved challenging without clear awareness of the underlying spatial relationships.

Together, these four scenarios represent common design states where nonvisual webpage authors may lose awareness of how content and layout interact.

\subsubsection{Procedure}
Co-design Session 2 was conducted on a different date from Co-design Session 1, with all three blind participants from the first session participating in this follow-up study.
Each session began with an explanation of the four scenarios and the study goals. 
Participants worked through each scenario while offering feedback on how the system should help them detect content–layout conflicts and decide how to resolve them.
One researcher role-played the system, providing audio feedback and clarifications, while another assisted with manipulating the tangible brackets when needed.
We observed participants' strategies, probed their reasoning with follow-up questions after each scenario, and concluded with an open discussion of preferred feedback types and timing. Each 90-minute session was video- and audio-recorded, transcribed, and anonymized for analysis.

\subsubsection{Findings}
We present findings from Co-design Session 2. We continue the numbering from Session~1.

\begin{figure}[t]
  \centering
  \includegraphics[width=\columnwidth]{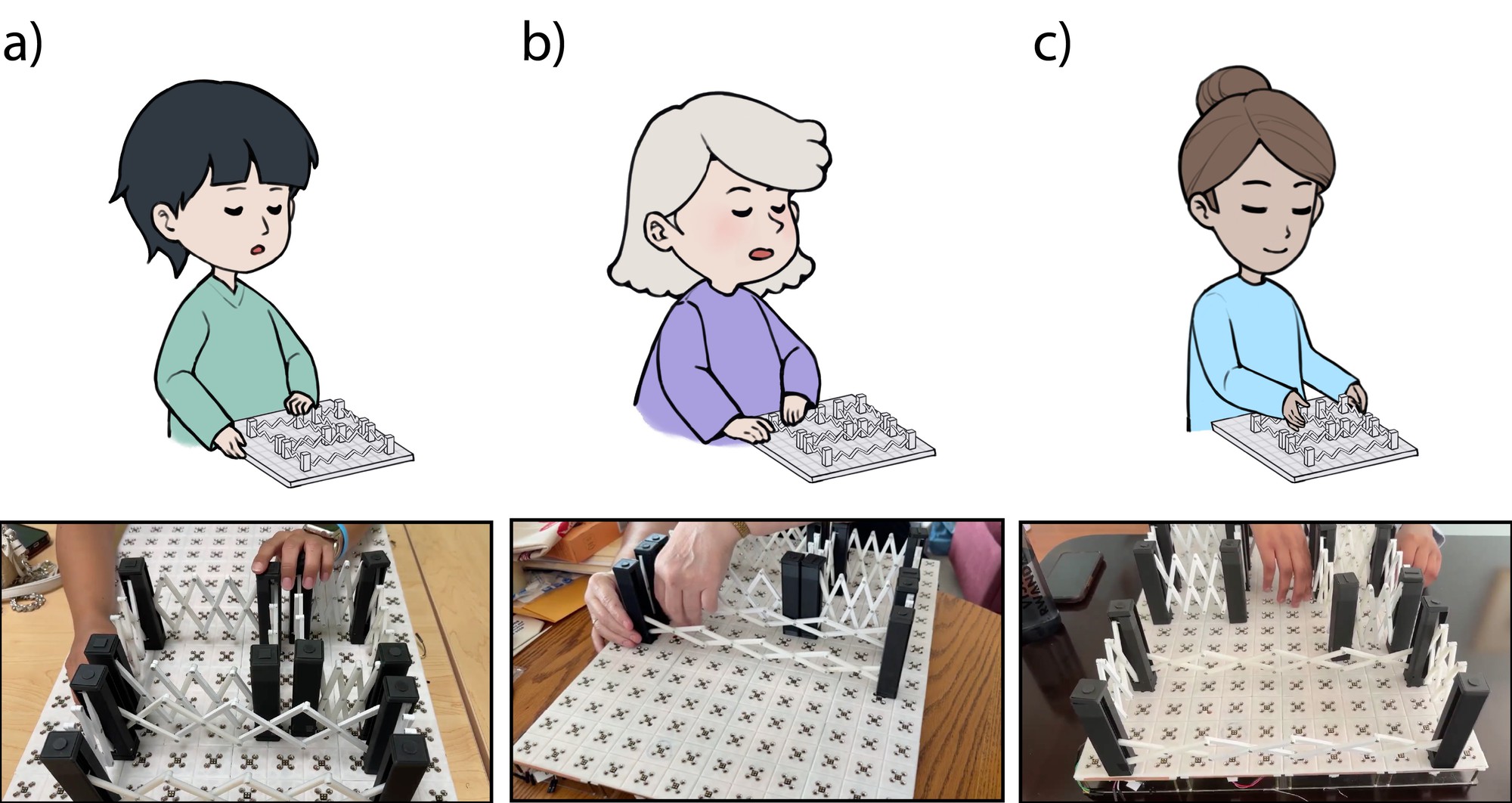}
\caption{Examples of three touching behaviors. a) P1 used touch to identify the category of a bracket by feeling the tactile patterns on its top. b) P2 used their hands to touch brackets, the baseboard, and tactile patterns to check alignment and gauge size. c) P3 swept the board and used their hands to touch multiple brackets to recall layout decisions.}
  \Description{Three photos show different tactile interaction behaviors: (a) using touch to identify the category of a bracket by feeling the tactile patterns on its top when they lost track of its category, (b) touching brackets, the baseboard, and tactile patterns to check alignment and size, and (c) sweeping and touching multiple brackets to recall layout decisions.}
  \label{fig:sweeping_board}
\end{figure}

\textbf{F2. Hands as the primary channel for monitoring, confirming, and planning.}
Across all scenarios, participants relied on their hands as the primary channel for understanding and reviewing their webpage designs (Figure ~\ref{fig:sweeping_board}). 
For example, when participants lost track of the category of a bracket, they used touch to perceive the tactile patterns on the top of the bracket to remind themselves of its category. When they needed to check alignment, gauge the size of a bracket, or simply recall their layout decisions, they naturally used their hands to touch the bracket or trace the grooves on the baseboard. Participants also swept their hands across the design canvas to refresh their mental model of the layout and keep track of their overall design status. The grooves and tactile features embedded in the tangible interface helped them stay aware of the layout and prompted them to think about the corresponding content.

Hands were equally important for planning future actions. Participants frequently kept a hand on a bracket while deciding on their next steps. When referring to a specific web element, they always placed a hand on that bracket or pointed to it to ensure they were talking about the correct element on the design canvas. After placing a bracket, they kept a hand on it while confirming whether it was operating in the content input mode and able to accept the intended content. 

This reliance on hands for both monitoring and action underscores the need for systems that support hand-based interaction as a primary channel. At the same time, systems should clearly signal the current interaction mode and help disambiguate user intent so that touch-based exploration does not lead to confusion or unintended actions.

\textbf{F3. Preference for actionable audio feedback on content-layout fit.}
When hands were not sufficient to identify or resolve design challenge such as content-layout misfit, participants turned to audio support. However, most blind participants did not simply ask for “more description.” Instead, they articulated specific ways audio feedback should reveal the content–layout relationships and help them evaluate whether their design choices achieved the intended effect.

When entering text, participants may struggled to judge whether the content would fit within the bracket’s available space.
They asked questions such as \textit{``How many words did I insert?''} and \textit{``How many words can this bracket hold?''}
In one text-entry instance, P2 paused mid-sentence to count words and reconsider their phrasing, unsure whether the text would fit within the bracket. 
She explained, \textit{“I'm quiet because I'm just counting the words (in the text content, and thinking about) how I want to express this.”} 
P1 similarly asked whether touching a bracket could report the number of words it contained. These behaviors suggest that understanding content capacity is cognitively demanding without explicit feedback. Participants wanted the system to provide bracket capacity and remaining space so they could decide whether to edit the text or adjust the layout rather than rely on their own memory or guesswork.

Participants described similar challenges with media elements. Choosing an image and inserting it was accessible with the keyboard, but understanding how the image was displaced within the bracket was not. They requested information about orientation, aspect ratio, and the amount of whitespace surrounding the image. P1 explained, \textit{``It does matter if the picture is vertical or landscape. I don’t know how the picture views in the bracket unless I `see' it.''} 
Participants also wanted the system to suggest concrete adjustment options in terms of rows and columns they could physically trace with their hands. As P3 noted, \textit{``I want it (the system) to tell me and to recommend what I should do, if I should extend it by width or height, because I really don’t want whitespace in my image bracket.''}

Participants also asked for feedback that summarized the design at a broader level. While they could touch individual brackets to recall type and content, they still struggled to keep track of empty regions or gaps on the canvas. P2 explained that they wanted feedback on where unused space remained so they could avoid leaving \textit{“too much whitespace,”} which would reduce visual appeal.

Taken together, these observations indicate that audio feedback should provide actionable information that goes beyond simple descriptions. Systems need to report how content fits within available space, highlight spatial relationships, and offer concrete options for resolving mismatches so that blind designers can make informed decisions about their layouts.

\subsection{Design Opportunities}
Drawing on findings F1–F3 from the two rounds of co-design, we identified three opportunities that directly address how blind individuals coordinate content and layout, use touch as a multifaceted interaction channel, and rely on actionable guidance to manage growing design complexity. These opportunities guided the development of the \tangiblesite system.

\textbf{D1. Maintain a persistent, up-to-date representation of the evolving page.} Participants frequently shifted between content-first and layout-first strategies, but this flexibility became difficult to sustain as more elements were added. To support content and layout co-evolving during design, tools should maintain an updated model of all content and layout information and make~it~accessible~on~demand. 

\textbf{D2. Provide immediate, touch-aligned feedback that clarifies system state and user intent.}
Touch served many roles at once: inspecting past decisions, confirming current actions, and planning future edits. Systems should respond to touch with localized haptic cues that confirm selection, mode, and the status of the element being handled. Feedback needs to be sensitive to the context of touch so that review, editing, and exploration can stay risk-free and predictable.

\textbf{D3. Offer actionable audio feedback about content–layout relationships.}
Participants did not simply want a description. They wanted feedback that explained how content fit inside containers, revealed mismatches in image orientation or text capacity, and suggested concrete adjustments they could choose from. Tools should provide concise structural audio guidance, including state summaries and recommendations for resolving conflicts across elements. This support scales from single-element edits to multi-element layouts, where participants struggled most to maintain a coherent mental model.

\section{\tangiblesite}
\begin{figure*}[t]
    \centering
    \includegraphics[width=\textwidth]{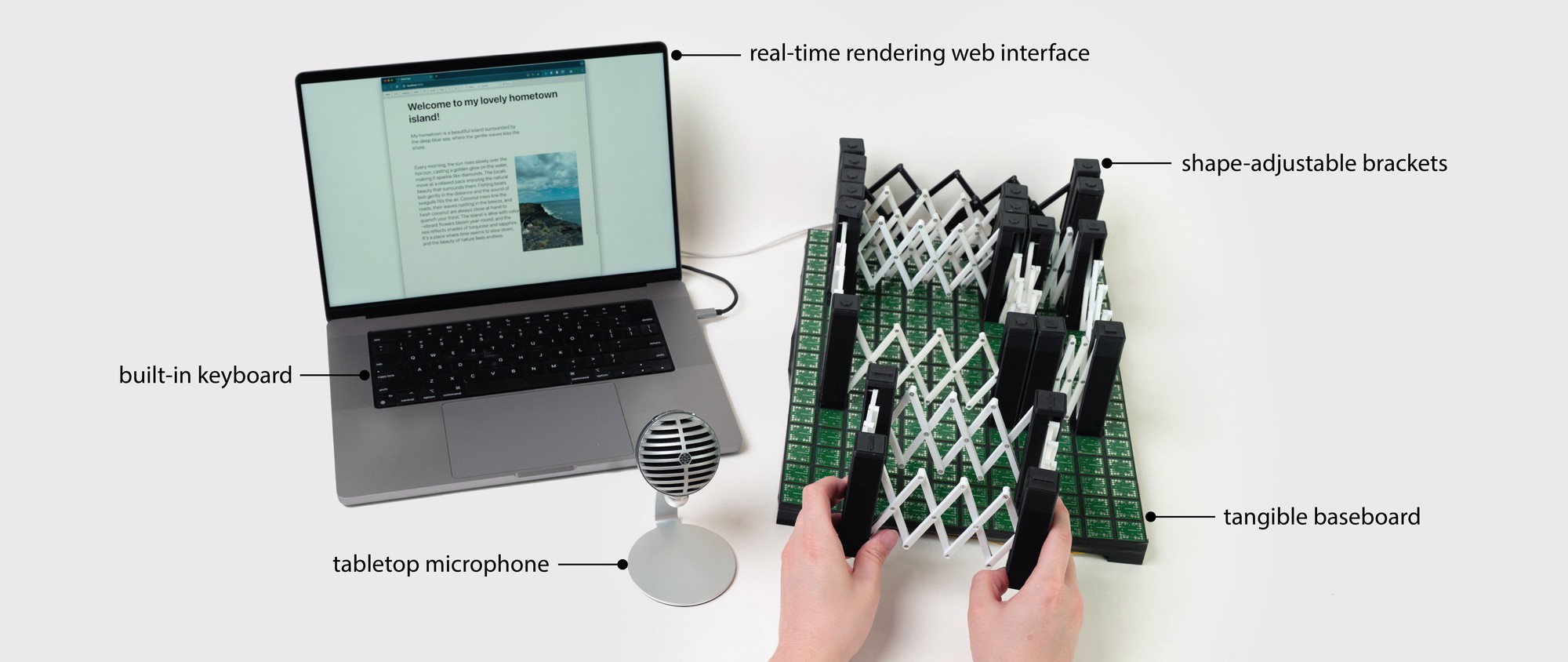}
    \caption{System overview. \tangiblesite supports accessible webpage authoring by coupling tangible layout manipulation with real-time multimodal feedback. Shape-adjustable brackets on a baseboard represent webpage elements and preserve structure across edits. Touch and audio feedback operate at both element and page levels, supporting inspection, verification, and overall layout understanding. A tabletop microphone enables voice commands, and a built-in keyboard supports text entry.}
    \Description{Photo of the \tangiblesite setup on a table. On the left, a hand manipulates a grid baseboard populated with upright pillars and touches brackets to select an element. On the right, a laptop displays a live webpage preview titled “Welcome to my lovely hometown island!” with a paragraph and two images. A small desktop microphone sits between the baseboard and the laptop, indicating voice input.}
    \label{fig:tangibleSite_overview}
\end{figure*}

\tangiblesite is an accessible web authoring tool for blind designers the joint development of content and layout. It maintains an up-to-date representation of the page’s structure and content, and provides real-time multimodal feedback via tangible, auditory, and speech-based interactions. Figure~\ref{fig:tangibleSite_overview} shows the overall system setup.

\subsection{System Walkthrough}
To illustrate how \tangiblesite supports nonvisual webpage design, we follow Zoe, a blind college student interested in web development with no prior coding experience. In everyday use, Zoe browses the web with a screen reader and enjoys reading blogs. With \tangiblesite, she sets out to create her first blog page from scratch on her own.

\tangiblesite comprises a tangible baseboard and a set of shape-adjustable brackets that represent the webpage canvas and elements, along with a companion browser interface that supports content entry (via voice or keyboard) and provides audio feedback. The browser interface communicates with the tangible baseboard to track Zoe’s actions, detect design issues, save content, render a live page preview, and maintain an up-to-date representation of the evolving page.

\textbf{Place elements and receive immediate state confirmation.}
Zoe wants to write an introduction to her hometown, with short text paragraphs and a few photos and videos. She begins by creating a title. Zoe selects a text bracket and places it at the top of the baseboard. Magnets embedded in both the baseboard and bracket provide snap-to-grid alignment. The system detects the bracket’s type, location, and size, announces a spoken summary (e.g., \textit{``Text bracket detected, size 2 by 8, location at row 1 and column 3.''}), and renders the corresponding container in the browser interface.

\textbf{Enter content via voice or keyboard.}
To enter the title, Zoe touches the bracket and feels a vibration confirmation indicating that it is selected and ready for content entry. This confirmation clarifies when an element is selected for subsequent commands, reducing accidental edits or content entry on the wrong element. She then says the wake word followed by the ``Title'' command, dictates the title, and says ``Stop'' to end dictation. Later, when adding media, Zoe uses the browser interface with her screen reader and keyboard to select files for upload.

\textbf{Iterate content and layout to resolve misfit.}
After Zoe enters the title, the system checks whether it fits inside the bracket and reports a mismatch (e.g., \textit{``Text exceeds bracket capacity. Consider shortening the text or expanding the bracket.''}). Zoe enlarges the bracket by expanding it one column to the right. The system recognizes the resized bracket as the same element and confirms that the title now fits, preserving continuity as she iterates between content and layout. Zoe then adds a second text bracket for a short paragraph; after similar steps, the system announces element details and fit status to support iteration.

With the text in place, Zoe adds a photo. She picks up an image bracket, places it below the previous text bracket, and left-aligns it with the text. The system senses the bracket’s type, size, and location and provides a spoken summary. Zoe selects the bracket, issues the ``Media'' command, and a browser dialog opens for file selection. Using her screen reader and keyboard, she navigates to an image stored on her computer and uploads it. The system renders a live preview. If it detects a mismatch between the image orientation and the bracket dimensions, it provides an actionable recommendation (e.g., \textit{``Image inserted; consider narrowing the bracket to reduce whitespace.''}). Zoe follows the suggestion by repositioning the image bracket with a narrower column width.

\textbf{Review the evolving page and run checks on demand.}
Zoe repeats this workflow to add multiple text, image, and video brackets. When she forgets what an element contains, she touches the corresponding bracket and requests a brief element summary. The system reads the element’s type and content status, helping her reorient without reconstructing the page from memory.

Periodically, she issues the ``Check'' command, which reads aloud a structured layout summary, including the number and types of brackets on the baseboard and which grid regions are empty or occupied. The system flags issues (e.g., text overflow or media misalignment) and proposes fixes, reducing the cost of reorientation as the design grows. Zoe alternates between editing content and reshaping or repositioning brackets, allowing content and layout to co-evolve. After several content iterations and layout changes, Zoe is satisfied with her webpage and feels comfortable sharing it with her friends.

\subsection{Interaction Design}

The interaction design of \tangiblesite is grounded in the design opportunities distilled from our co-design sessions: the need for a persistent representation of the evolving layout, a way to distinguish user intentions during touch, and clear guidance when content and layout fall out of alignment. \tangiblesite integrates tangible, haptic, and speech-based modalities to support these needs. 

\textbf{Tangible interactions for maintaining spatial grounding (D1).}
During co-design, participants depended on stable spatial cues to rebuild their mental model of the page, especially once multiple elements were present. \tangiblesite retains the familiar tactile grid and snap-in brackets introduced in prior work, but extends their function by linking every physical bracket to a persistent digital record of its type, size, and associated content. When a bracket is moved, resized, or temporarily lifted from the board, the system preserves its content and layout metadata, allowing users to restore or modify the element without losing track of what it represents. This integration of tactile structure with a persistent layout memory supports D1 by giving blind designers a reliable spatial anchor they can return to at any time, without having to reconstruct the entire page from memory.

\textbf{Touch and haptic cues for clarifying system state (D2).}
Participants often touched brackets for different purposes: to review what an element contained, to confirm whether the system recognized their selection, or to prepare the next action. Since these intentions are indistinguishable from the system’s perspective without additional cues, \tangiblesite introduces context-aware haptic feedback that disambiguates these touch states. A brief vibration confirms that a bracket has been selected and that subsequent speech commands will be routed to it. A long press triggers a state readback that announces the bracket’s type, size, and any stored content or media. These immediate, touch-aligned cues reduce uncertainty about whether the system registered an action and help users maintain orientation as the layout changes. This design directly supports D2 by ensuring that tactile exploration and manipulation are always paired with clear system acknowledgment.

\textbf{Audio guidance for understanding content–layout relationships (D3).}
Co-design participants did not want more description. Instead, they wanted information that helped them decide what to do next. To support this need, \tangiblesite provides audio feedback that explains how content and layout interact, paired with concrete adjustment options. 

When text is entered, the system checks fit and reports the information exactly as implemented, for example: \textit{``The current number of characters inside the bracket is 20. The maximum is 15. The recommended number is 10.''} This allows users to determine whether to revise wording or resize the bracket.

\begin{figure*}[!b]
    \centering
    \includegraphics[width=\textwidth]{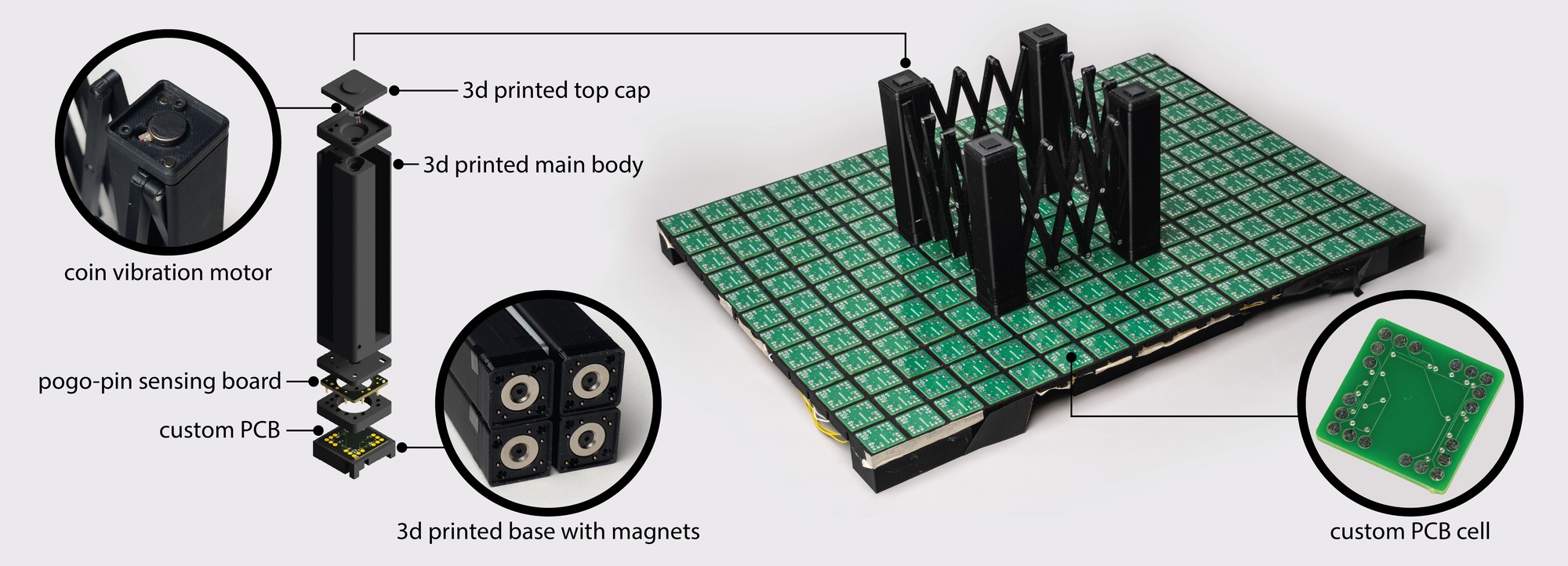}
    \caption{\tangiblesite hardware components. 
    The baseboard detects bracket type, location, and size using custom PCB cells. An exploded view shows the modular pillar design with magnetic attachment, electrical sensing, and vibrotactile feedback.
    }
    \Description{A figure of \tangiblesite hardware. A bracket is placed on the tangible baseboard; the board is equipped with custom-made PCBs that can detect bracket type, location, and size. On the left, a exploded view of one pillar, bottom-to-top: 3D printed base, custom PCB, spring pogo-pin sensing board, 3D-printed main body, cavity for a coin vibration motor, and a top cap with a tactile pattern.}
    \label{fig:tangiblesite_system_combine}
\end{figure*}

When images or videos are inserted using the ``Media'' command, \tangiblesite opens the folder containing all the pre-prepared media files, enabling navigation with standard screen reader tools. After upload, the system evaluates the media–container relationship and verbalizes an actionable recommendation, such as \textit{``Image inserted; one column remains empty on the left and right. Consider narrowing the bracket by one column on each side.''}

Speech commands are organized around common authoring tasks. Commands such as ``Title'', ``Text'', ``Next line'', and ``Stop'' control content entry. The ``Check'' command retrieves a summary of the current webpage state: the system verbalizes the total number of each bracket type on the baseboard and calculates the percentage of whitespace not occupied by brackets.
This summary highlights unresolved overflow, underflow, or spatial gaps without requiring users to sweep the board manually.

Together, these verbal cues turn audio feedback into a decision-making resource. Rather than passively narrating the layout, \tangiblesite explains fit, points out mismatches, and suggests next steps. This realizes D3 by giving blind designers the information and options they need to iterate with intention and resolve content–layout conflicts nonvisually.

\textbf{Supplementary keyboard input for precise text refinement (D3).}
Although speech-based input is the primary method for entering and revising content, \tangiblesite also supports conventional keyboard interaction. This option gives blind designers greater control over fine-grained edits, such as selecting, deleting, or rearranging text. Many blind individuals already work efficiently with screen reader software and familiar shortcuts like “Ctrl + C” and “Ctrl + V,” so the keyboard serves as a natural extension of the system’s audio-guided content editing. This complements the actionable feedback in D3 by enabling text refinement that dictation alone may not provide.

\subsection{System Implementation}
\subsubsection{Hardware}
As shown in Figure \ref{fig:tangiblesite_system_combine}, the baseboard and tangible brackets of our system are inspired by TangibleGrid~\cite{li_tangibleGrid}, but incorporate several key design differences. The baseboard measures 305 mm $\times$ 406 mm and contains a grid of 192 cells. Each cell serves two functions by physically aligning brackets placed on top and electronically connecting to them for sensing and vibration. To support these functions, each cell in \tangiblesite has a layered structure, consisting of a 3D-printed holder at the base, a bolted countersunk ring magnet (15 mm diameter, 3 mm thickness), and a custom-designed PCB at the very top. The PCB contains 20 contact points arranged in central symmetry, together with the ring-shaped magnet, ensuring that brackets connect correctly to the baseboard regardless of orientation. The circuit detects bracket touch and activates vibration feedback when required.

The tangible bracket, shown in Figure~\ref{fig:tangiblesite_system_combine}, consists of four pillars connected by scissored linkages. While its appearance resembles that of TangibleGrid, the implementation differs substantially.
Each pillar is designed as a multi-section structure to support both direct touch and localized vibration feedback. At the top, a 3D-printed tactile pattern made from conductive PLA filament enables capacitive touch sensing. Directly beneath the pattern, a vibration motor is embedded to provide vibrotactile feedback, with its pins routed through the pillar to a custom PCB concealed at the base. The PCB incorporates five spring-loaded pogo pins that interface with the physical baseboard, along with a surface-mount resistor that allows the \tangiblesite system to automatically identify the bracket type, compute its size, and register touch interactions. Finally, a ring magnet matching that of the baseboard is bolted to the very bottom of the pillar, flush with the pogo pins, to ensure secure physical alignment at all times.

The size of the bracket is adjustable, ranging from 50 mm $\times$ 50 mm to 300 mm $\times$ 300 mm. The maximum width of 300 mm allows the bracket to cover the entire width of the web canvas, which is useful when users want to design full-width elements, such as banners. The height of the bracket is fixed at 112 mm.

\subsubsection{Software}
Our software consists of three main components: (1) the front end, which renders the webpage and provides audio feedback; (2) the back end, which communicates with the \tangiblesite hardware to receive information about the brackets; and (3) the database, which stores the web layout design and content. The front end is built with the React framework and TypeScript. We utilize the SpeechRecognition Web Speech API for voice recognition and the SpeechSynthesis Web Speech API for audio feedback. The front end is supported by a back end server running on Express, a minimal web framework in Python. The server receives all web layout design information, such as bracket size and location, from the \tangiblesite hardware and stores it in the database.

\begin{figure*}[!t]
  \centering
  \includegraphics[width=\textwidth]{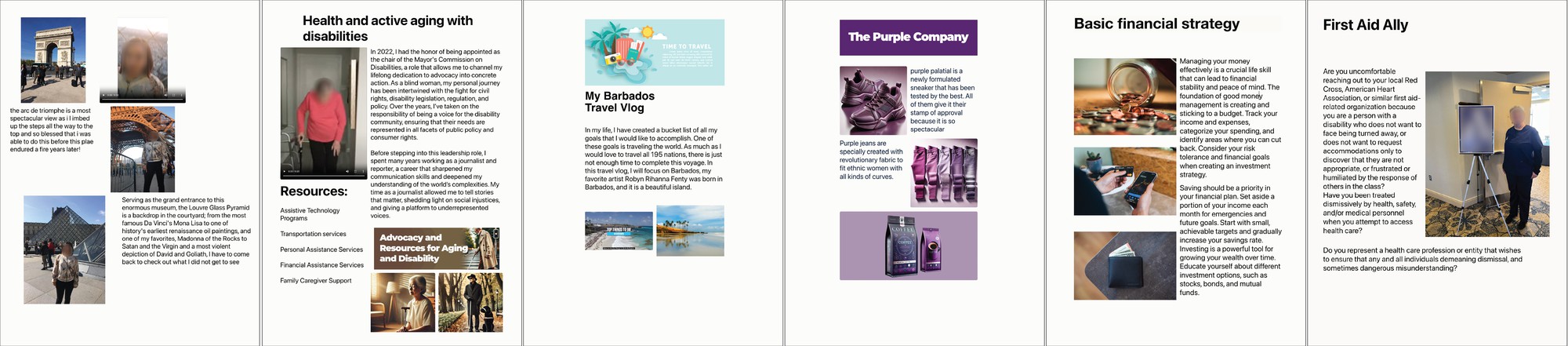}
  \caption{Six webpages designed by blind participants with \tangiblesite during the evaluation study.}
  \Description{Six webpages produced by blind participants in the evaluation study using the tangible prototype; Designs include: a) a travel diary with descriptive text and multiple photos, b) an article titled “Health and active aging with disabilities” with a left Resources list, c) a blog titled “My Barbados Travel Vlog” with a hero banner and image gallery, d) a product promo titled “The Purple Company” with stacked product shots and ad copy, e) an advice page titled “Basic financial strategy” with supporting images, and f) an advocacy page titled “First Aid Ally” with long-form text and a photo.}
  \label{fig:evaluation_rendering}
\end{figure*}

\section{User Evaluation}
We conducted a formative user study to evaluate the effectiveness of \tangiblesite in enabling blind individuals to create and iterate on their own webpage.

\subsection{Participants}
We recruited a total of six participants, including five females and one male. 
Three took part in the co-design study, while the other three participants were newly recruited through a local NFB email group.
Among participants who disclosed their age, ages ranged from 19 to 69 years (\textit{M} = 45.4, \textit{SD} = 16.6); one participant chose not to disclose their age. All participants self-reported being blind.

The newly recruited participants (P4--P6) brought varied levels of web design experience. P4 had no prior experience with web design. P6 had some familiarity with WordPress but was only comfortable editing text content. P5 had extensive web design experience and was familiar with website template platforms such as WordPress and Drupal. They hired someone to initially create their personal website and now maintain it using HTML, CSS, and PHP.

\subsection{Procedure}
The study had two sessions: a practice session and an evaluation session. In the practice session, participants familiarized themselves with \tangiblesite through guided tasks such as creating and deleting webpage elements and requesting audio feedback about individual elements or the overall page layout. Participants were also encouraged to freely explore and practice both layout design and content entry.

Once participants became familiar with the system, they proceeded to the evaluation session, where they were tasked with designing a webpage based on one of the following themes: a portfolio, personal blog, travel diary, or small business homepage. To accommodate time constraints, participants were encouraged to brainstorm initial design concepts and prepare content in advance of the study visit. If needed, we also offered the option to generate content using ChatGPT during the session. Upon completing the evaluation session, participants took part in a post-study interview, which included open-ended questions and Likert-scale ratings to gather feedback on their experience using the \tangiblesite. 

The entire study lasted approximately two hours. All participants were compensated at a rate of \$35 per hour. The study was approved by the university's IRB.

\subsection{Study Results}
\begin{figure*}[!b]
  \centering
  \includegraphics[width=\textwidth]{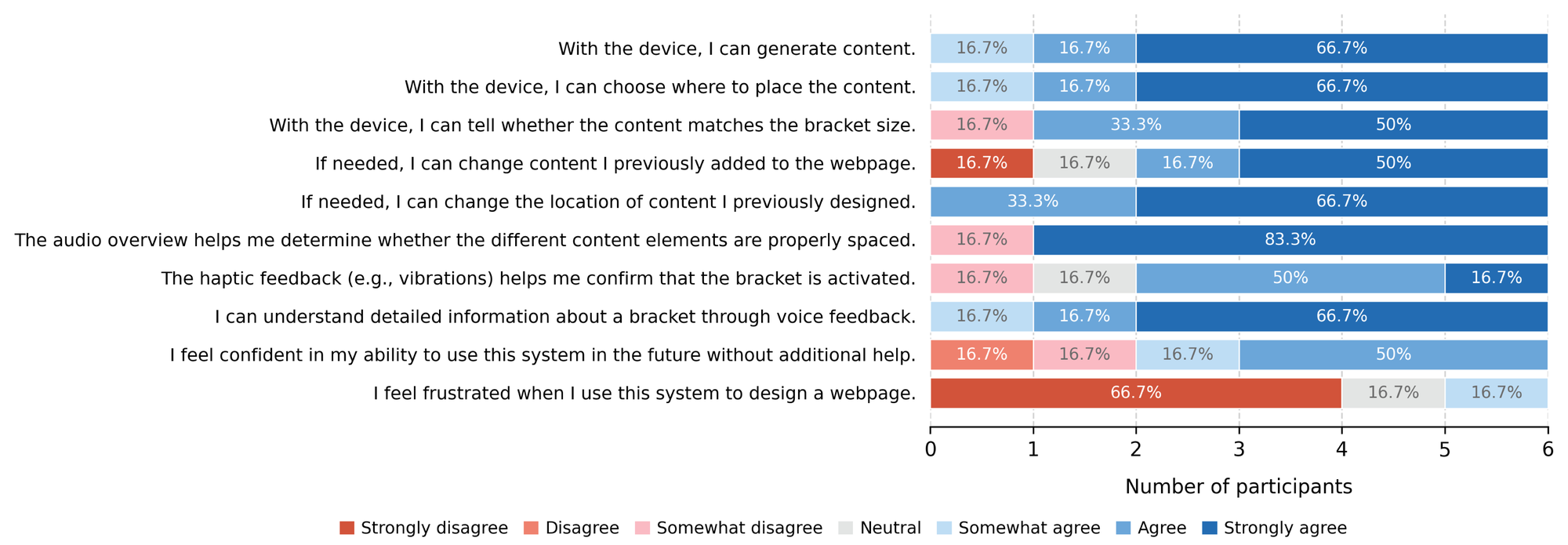}
  \caption{Participant self-reported ratings for ten Likert scale statements about web design experience with \tangiblesite.}
  \Description{Horizontal stacked bar chart with ten rows (self-subjective rating questions) and an x-axis labeled “Number of participants, 0–6.” Each bar is subdivided into the seven Likert choices (from Strongly Disagree to Strongly Agree). Most bars are dominated by the agree end of the scale. Notable values visible on the bars include 66.7\% Strongly agree for “I can generate content” and “I can decide on the location,” 50\%–66.7\% agreement that content fits the bracket and can be changed, 83.3\% agreement that the audio overview helps with spacing, and 66.7\% Strongly disagree for “I feel frustrated when I try to use this system}
  \label{fig:likert}
\end{figure*}

All six participants completed a webpage during the evaluation session (Figure~\ref{fig:evaluation_rendering}). Likert ratings in Figure~\ref{fig:likert} indicate that participants were able to use \tangiblesite to generate content, design webpage layouts, and iteratively revise web elements. Most participants noted that the combination of interaction modalities helped them understand both the overall webpage structure and the details within individual elements. Participants expressed enthusiasm about the design possibilities enabled by \tangiblesite, noting that it allowed them to accomplish tasks they had rarely considered feasible. Below, we organize findings by how D1--D3 manifested in independent authoring.

\textbf{Persistent spatial grounding supported diverse iterative workflows (D1).}
Participants used the physical layout and in-system content memory to manage complexity and revisit earlier decisions. Those who preferred sequential workflows (P3, P4) finalized elements one at a time, relying on the system’s representation and feedback about both structure and content. 

\begin{quote}
\textit{“You know there are so many things to keep track of, and I forget things so I like to finalize each element before I move to the next.” -- P3}
\end{quote}

Others leveraged D1 to support more flexible, global reasoning. P2 arranged all three image brackets before inserting photos, explaining that they wanted to understand spacing and size first. P4 experimented with relocating a text bracket to balance future photos across the page; when the system reported that the new location could not hold the original text, they restored the previous layout. Together, these behaviors show how persistent spatial grounding enabled exploration and reduced the risk of losing track of content or creating layout-content inconsistencies.

\textbf{Touch-aligned confirmation increased confidence and reduced errors (D2).}
Participants used touch both to understand the layout and to verify selection and system status. Many described vibration feedback as essential for maintaining confidence during manipulation. P2, for example, emphasized that vibration provided immediate confirmation that the intended bracket was selected, increasing their sense of control during the design process. Participants also used touch-triggered readbacks to reorient themselves, reducing reliance on memory as layouts grew more complex. Overall, vibration confirmation and touch-triggered readbacks reduced ambiguity and helped participants act on the intended element with confidence.

\textbf{Actionable audio guidance enabled informed refinements to content and layout (D3).}
Real-time analysis of bracket content, spacing, and fit allowed participants to identify and correct issues they would otherwise miss. P1 repeatedly adjusted bracket sizes until all text fit, explicitly crediting the system’s feedback:

\begin{quote}
\textit{“It's telling me how many characters belong in a specific box... So it's giving me the necessary information that you need without overkill.” -- P1}
\end{quote}

\begin{figure}[!t]
  \centering
  \includegraphics[width=\columnwidth]{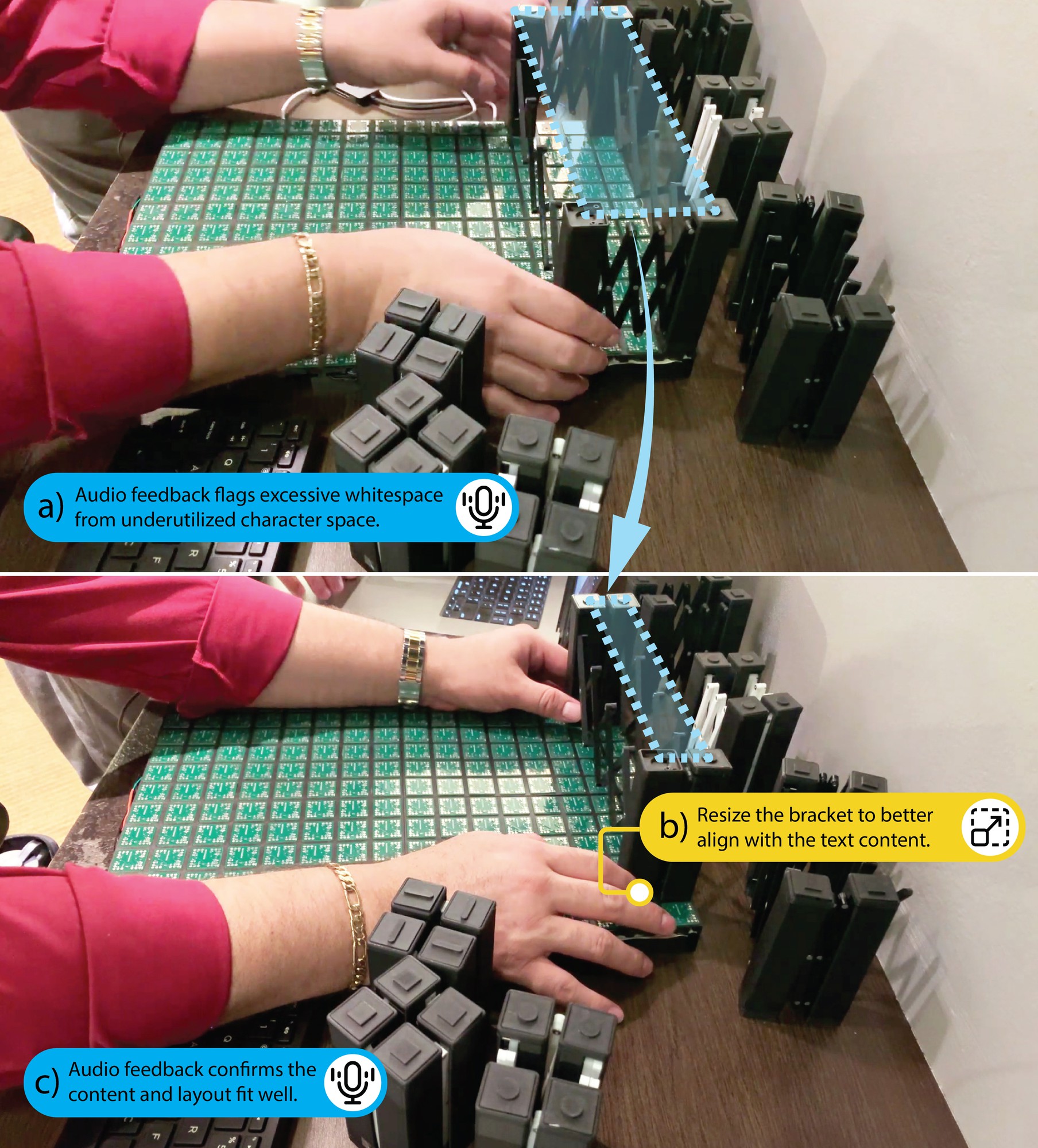}
  \caption{Iterative design refinements made during the evaluation session: The participant receives a) system-generated audio feedback detailing the current status of the text bracket. Based on this feedback, the participant decides to b) resize the bracket to better accommodate the content, continuing until c) the audio feedback confirms that both the content and layout fit appropriately.}
  \Description{Two photos are included in this picture. They were taken from an evaluation study showing blind participants made design iterations with TangibleSite. a) The participant receives system-generated audio feedback detailing the current status of the text bracket. b) Based on this feedback, the participant decides to resize the text bracket to better accommodate the content.}
  \label{fig:eva_adjustment}
\end{figure}

Similarly, when inserting images, participants responded to feedback about unused space or aspect-ratio mismatches by resizing brackets. For example, P5, who had web design experience, noted that combined tactile and audio feedback improved awareness of alignment and whitespace, which can be difficult to perceive nonvisually (Figure~\ref{fig:eva_adjustment}).

Participants frequently used the “Check” command to receive an overview of the entire page, learn which regions were empty or occupied, and identify potential misalignment. Together with element-level recommendations, these page-level summaries supported both local fixes and global refinement.

\section{Discussions}
\subsection{Multimodal Balance: Supporting Flexible Design without Overload}

Our co-design findings showed that blind participants prefer to approach webpage creation as a fluid process in which content and layout co-evolve. \tangiblesite supports this flexibility with persistent system memory (D1), immediate touch-aligned confirmation (D2), and actionable audio guidance (D3). However, this flexibility can still become fragile as design complexity increases. As multiple elements accumulated on the canvas, participants could still feel burdened by remembering prior decisions, monitoring content fit, and keeping track of spatial relationships.

These observations highlight an inherent tension. Providing full control over both content and layout increases creative freedom but also increases cognitive effort, especially for blind individuals who are new to web design. Several evaluation participants noted that they needed time to learn and recall commands and determine next steps, and relied heavily on system prompts during early use. This suggests that accessibility is not only a matter of adding features or additional modalities, but also of managing how much information and how many choices the system presents at any moment.

To address this tension, future versions of \tangiblesite could incorporate adjustable guidance levels. For novices, the system could reduce decision load through stepwise workflows (e.g., templates or guided build-up modes), limit available actions to those most relevant in context, and prioritize high-impact feedback such as overflow, misalignment, and spacing conflicts. The system could also apply constraints or offer auto-adjustments when layout limits are reached. Such scaffolding could help blind designers build an initial mental model of content--layout relationships while still supporting iterative refinement.

For experienced designers, \tangiblesite could provide richer but more compact summaries and more direct control, including intent-aware suggestions inferred from element type and structure (e.g., headings, bullet lists, banners, and image/video sections), to support complex designs without interrupting flow. Overall, adapting guidance to user expertise may preserve flexible content-layout iteration while reducing the likelihood of overload.

\subsection{Opportunities for Adaptive and Semi-Automated Layout Support}
The evaluation sessions showed that participants could reliably add, modify, and inspect individual elements with \tangiblesite. However, interactions involving many elements or large structural changes required significant manual effort. 
Participants often needed to lift and reposition multiple brackets, recheck content fit, and reestablish their mental model of the layout after each change. Even with strong physical affordances and system feedback, these actions can be taxing when repeated.

These observations suggest an opportunity for partial automation that preserves agency while reducing mechanical effort, for example through motorized actuation of the shape-adjustable brackets (or mechanisms beneath the baseboard that can reposition them). With explicit approval, the system could automate common operations such as aligning elements, snapping to the grid, redistributing spacing, or instantiating frequently used layout templates. Consistent with our earlier discussion of adaptive feedback, the system could offer multiple levels of assistance, ranging from lightweight alignment aids to higher-level template generation to accommodate different levels of web design experience.

Several participants expressed interest in recognizable layout patterns (e.g., blogs, portfolios, small business sites). Motorized action could allow the system to lay down a template directly on the baseboard, preserving the tactile grounding that supports D1 and D2 while reducing setup time. For novice designers, templates could provide an initial scaffold for understanding content-layout relationships and support focused content creation within a stable structure, potentially with optional auto-fit behaviors (e.g., resizing to avoid overflow). For more experienced designers, motorized reconfiguration could support complex structures that are difficult to realize nonvisually, such as floating navigation bars or pop-up windows, and could extend to responsive variants that adapt across device breakpoints.

Importantly, any automation should follow a preview-and-accept interaction pattern. The system can propose an adjustment based on goals or content constraints, but changes should be confirmed through touch and verification of the resulting physical configuration. This approach may reduce repetitive effort while maintaining control and transparency in the design process.

\subsection{Design Advanced Webpages with AI}
While \tangiblesite currently supports basic design elements (text, image, video), modern webpages are typically far more complex, incorporating forms, animations, and dynamic layouts that are especially challenging for nonvisual design.
Recent advances in generative AI make it possible to create more complex webpages through natural language prompts, suggesting a role for AI as a design assistant integrated throughout the \tangiblesite workflow.

Prior work demonstrates AI capabilities such as proposing text, generating images~\cite{genassist_2023mina}, and editing video clips~\cite{avscript_2023huh}, which could help blind designers move past blank-page setup. AI could also translate high-level styling intent (e.g., theme, background color, contrast) into CSS/HTML, enabling appearance adjustments without writing code. Beyond styling, AI could synthesize interactive components or small programs, including simple games, and embed them within webpages to support richer interactions. Integrated into \tangiblesite, these capabilities could shift effort away from low-level implementation toward composition and higher-level design decisions.

\subsection{Supporting Collaborative Design}
While \tangiblesite is currently oriented toward independent use, we see its potential as a bridge between blind and sighted collaborators. Blind designers gain tactile and audio insight into structure and content, while sighted collaborators can view the visual rendering. Real-time synchronization between tangible and digital representations could enable parallel editing and shared reasoning across modalities.

Extending \tangiblesite for such collaborative use would require features that support shared understanding and co-creation. For example, real-time, fine-grained synchronization between the tangible and visual interfaces could allow both collaborators to work in parallel on the same design. Additional features, such as webpage templates and annotation tools, may also be needed to support a shared workspace. We see this as a promising direction for future development.

\subsection{Limitations}
We close by noting limitations that bound the scope of this work.

\textbf{Limited layout expressiveness.}
\tangiblesite represents layout through fixed bounding boxes on a uniform grid, without margins, padding, or finer typographic hierarchy. This simplification limits the range of spatial relationships and styling decisions that can be authored with the current system.

\textbf{Scope and accessibility of generated webpages.}
\tangiblesite outputs semantically structured HTML that reflects the tangible layout and authored content, but it does not currently support responsive behaviors, advanced styling, or full visual design control. The current pipeline also does not automatically generate accessibility metadata such as alt-text and labels.

\textbf{Study duration and scope.}
Our evaluation involved single-session interactions with six blind participants. While the study demonstrates feasibility and highlights promising patterns, longer-term use may reveal different strategies, breakdowns, or~learning~trajectories.

\section{Conclusion}
We present \tangiblesite, a multimodal website authoring tool that supports blind individuals in creating and iterating both content and layout within a single, integrated system. Informed by co-design sessions with blind participants, \tangiblesite maintains an up-to-date representation of the page’s structure and authored content and provides multimodal feedback to support ongoing refinement. Our user study shows that \tangiblesite reduces barriers to web design and supports iterative workflows, enabling blind designers to exercise greater control and creativity.

\begin{acks}
This work was supported in part by the National Science Foundation under Grant No. 2229885 (NSF Institute for Trustworthy AI in Law and Society, TRAILS). Any opinions, findings, conclusions, or recommendations expressed in this material are those of the author(s) and do not necessarily reflect the views of the National Science Foundation.
An LLM service was used for grammar and text editing.
\end{acks}

\bibliographystyle{ACM-Reference-Format}
\bibliography{my-reference}

\appendix

\end{document}